\documentclass[12pt]{article}
\usepackage{graphics} 
\newcommand{\e}{{\mathrm{e}}}
\newcommand{\mrm}[1]{{\mathrm{#1}}}
\newcommand{\pma}[2]{\! \! \mbox{\tiny $\begin{array}{l} +#1 
\vspace*{-0.7ex} 
\\ -#2 \vspace*{0.7ex} \end{array}$ } } 
\newcommand{\gev}{{\;{\mathrm{GeV}}}}
\newcommand{\gevc}{{\;{\mathrm{GeV}}/c}}
\newcommand{\gevcsq}{{\;{\mathrm{GeV}}/c^{2}}}

\begin{document} 
%
\begin{center}
{\bf \LARGE One-, Two- and Three-Particle Distributions from 158$A$
GeV/$c$ Central Pb+Pb Collisions} 
\end{center}
\vspace*{0.7cm}
{\small \noindent
M.M.~Aggarwal,$^{1}$ A.L.S.~Angelis,$^{4}$
V.~Antonenko,$^{5}$ V.~Arefiev,$^{6}$ V.~Astakhov,$^{6}$
V.~Avdeitchikov,$^{6}$ T.C.~Awes,$^{7}$ P.V.K.S.~Baba,$^{8}$
S.K.~Badyal,$^{8}$ S.~Bathe,$^{9}$
% ...
} 
\\ \vspace*{-0.1cm}
\begin{center} (WA98 collaboration) \end{center} 
\vspace*{0.2cm}
%
{\small
{$^{1}$~University of Panjab, Chandigarh 160014, India} \\
{$^{2}$~University of Rajasthan, Jaipur 302004, Rajasthan, India} \\
{$^{3}$~Variable Energy Cyclotron Centre, Calcutta 700 064, India} \\
{$^{4}$~University of Geneva, CH-1211 Geneva 4,Switzerland} \\
{$^{5}$~RRC Kurchatov Institute, RU-123182 Moscow, Russia} \\
{$^{6}$~Joint Institute for Nuclear Research, RU-141980 Dubna, 
Russia} \\
{$^{7}$~Oak Ridge National Laboratory, Oak Ridge, Tennessee
37831-6372, USA} \\
{$^{8}$~University of Jammu, Jammu 180001, India} \\
{$^{9}$~University of M{\"u}nster, D-48149 M{\"u}nster, Germany} \\ 
{$^{10}$~SUBATECH, Ecole des Mines, Nantes, France} \\ 
{$^{11}$~Gesellschaft f{\"u}r Schwerionenforschung (GSI), D-64220
Darmstadt, Germany} \\ 
{$^{12}$~Universiteit Utrecht/NIKHEF, NL-3508 TA Utrecht, The
Netherlands} \\
{$^{13}$~Lund University, SE-221 00 Lund, Sweden} \\ 
{$^{14}$~University of Tsukuba, Ibaraki 305, Japan} \\
{$^{15}$~Nuclear Physics Institute, CZ-250 68 Rez, Czech Rep.} \\ 
{$^{16}$~KVI, University of Groningen, NL-9747 AA Groningen, The
Netherlands} \\ 
{$^{17}$~Institute for Nuclear Studies, 00-681 Warsaw, Poland} \\ 
{$^{18}$~MIT, Cambridge, MA 02139, USA} \\
{$^{19}$~Institute of Physics, 751-005 Bhubaneswar, India} \\
{$^{20}$~University of Tennessee, Knoxville, Tennessee 37966, USA} \\
}
%
\normalsize
%
\abstract{
Several hadronic observables have been studied 
in central 158$A$ GeV Pb+Pb collisions
using data measured by
the WA98 experiment at CERN:
single $\pi^-$ and $K^-$ production, as well as two- and three-pion
interferometry.
The Wiedemann-Heinz hydrodynamical model has been fitted to the pion
spectrum, giving an estimate of the temperature and transverse flow
velocity.
Bose-Einstein correlations between two identified $\pi^-$ have been
analysed as a function of $k_T$, using two different parameterizations.
The results indicate that the
source does not have a strictly boost invariant expansion or
spend time in a long-lived intermediate phase.
A comparison between data and a hydrodynamical based simulation shows
very good agreement for the radii parameters as a function of $k_T$.
The pion phase-space density at freeze-out has been measured
and agrees well with the Tom\'a{\u{s}}ik-Heinz model.
A large pion chemical potential close to the condensation limit of $m_\pi$
seems to be excluded.
The three-pion Bose-Einstein interferometry shows
a substantial contribution of the genuine three-pion correlation,
but not quite as large as expected for a fully chaotic and
symmetric source.
}
%
%
\section{Introduction}

The study of single particle distributions of particles produced in
heavy-ion collisions gives access to the degree of thermal and chemical
equilibrium at freeze-out and allows the determination of the parameters
of hydrodynamical expansion models of the source.

The spatio-temporal extension of the interaction region created in such
collisions is not directly observable, but the study of Bose-Einstein
interferometry between identical particles provides information on the
geometry and on the dynamical evolution of the particle emission sources.
In particular, the correlations between produced pions give access to the
size of the homogeneity region, to the duration of emission, and to
various
parameters characterizing the spatial extension of the fireball
\cite{ZAK}.
In addition, by combining information from the source size in
momentum-space obtained by interferometry, and from the momentum-space
density provided by the single particle distributions, an average
phase-space density at freeze-out can be calculated.

Compared to the two-particle correlation, the three-particle correlation
can provide additional information on the chaoticity and asymmetry of the
source emission \cite{ANDR,LORS,HEISEL}.
In particular, the three-pion interference produced by a fully chaotic
source is sensitive to the phase of the Fourier transform of the source
emission function and, hence, to the asymmetry of the source.

In this paper, we present the analysis of single particle production, and
of two- and three-pion interferometry measured in the WA98 experiment for
central 158$A$ GeV $^{208}$Pb+$^{208}$Pb collisions at the CERN SPS.
In addition, estimates of hydrodynamical expansion model parameters,
temperature and transverse flow velocity are
extracted from these results.

\section{Experimental setup and data processing}

\begin{figure}[hbtp]
 \begin{center}
 \hspace{-1.35cm}
  \rotatebox{270}{%
   \resizebox{0.38\textwidth}{!}{%
    \includegraphics*{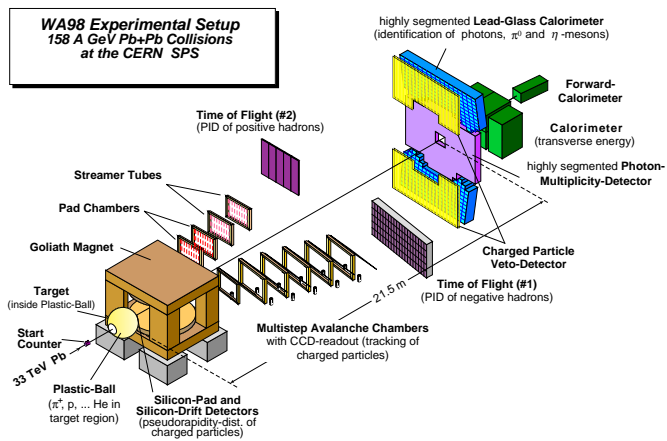}
   }
  }
  \caption{The WA98 experimental setup.}
  \label{fig:figure1}
 \end{center}
\end{figure}

\begin{figure}[hbtp]
\begin{center}
\hspace{-0.35cm}
\resizebox{0.5\textwidth}{!}{%
  \includegraphics{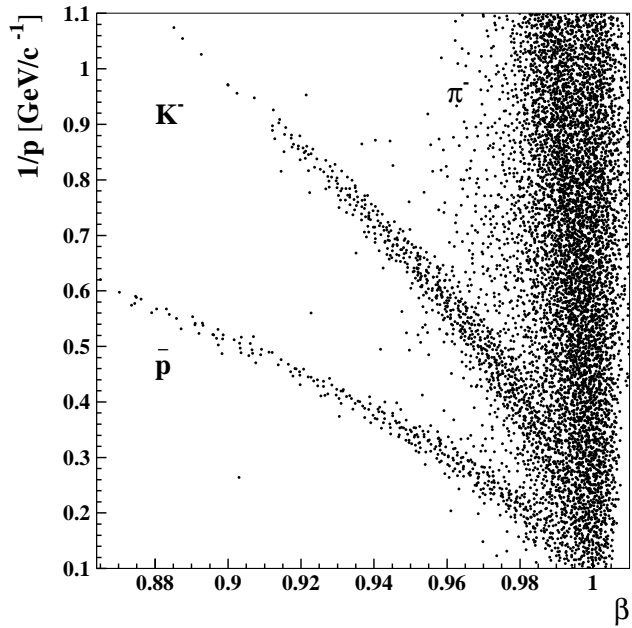}
}
\caption{Particle identification in the negative arm of the spectrometer.
}
\label{fig:pidbeta}
\end{center}
\end{figure}

The CERN SPS fixed target experiment WA98 \cite{PROP} combined large acceptance
photon detectors with a two arm charged particle tracking spectrometer.
The experimental layout is shown in Fig. \ref{fig:figure1}.
The 158$A$ GeV Pb beam interacted with a Pb target near the entrance
of a large dipole magnet.
The online trigger centrality selection used a forward calorimeter located
at zero degrees and a mid-rapidity calorimeter measuring the total
transverse energy in the pseudorapidity interval 3.2 $\leq$ $\eta$ $\leq$ 
5.4.
The results presented here have been obtained from an analysis of the
complete data set.
These data were taken with the most central triggers corresponding to
about 10\% of the minimum bias cross section of 6300 mb
\cite{PIZERO,agg00_}, with an average
of 330 participating nucleons per collision.
These quantities are estimated to have an overall systematic error of less
than 10\%.
The charged particle spectrometer made use of a 1.6 Tm dipole magnet with
a 2.4$\times$1.6 m$^2$ air gap which deflected the charged
particles in the horizontal plane into two tracking arms located
downstream, one on each side of the beam axis.
The results shown here are measurements of $\pi^-$ and $K^-$ observed in 
the negative particle tracking arm of the spectrometer.
This tracking arm consisted of six multistep avalanche chambers with
optical readout \cite{LUX}.
Inside the chambers, triethylamine (TEA) photoemissive vapour
produced UV photons 
along the path of
charged particles, these photons being subsequently 
converted to
visible light via wavelength shifter plates.
On exit, the light was reflected by mirrors at 45$^{\circ}$ to CCD
cameras
equipped with two image intensifiers.
The active area of the first chamber was 1.2$\times$0.8 m$^2$ and that of
the
other five 1.6$\times$1.2 m$^2$. Each CCD camera pixel viewed a region of
about 3.1$\times$3.1 mm$^2$ on the chambers.
Downstream of the chambers, at a distance of 16.5 m from the target, a
4$\times$1.9 m$^2$ time of flight wall
allowed for particle identification with a time resolution of better than
120 ps.
The resulting particle separation is shown in Fig.\ref{fig:pidbeta}.
The $\pi^-$ rapidity acceptance ranged
from $y=2.1$ to 3.1 with a rapidity average at 2.7, close to the
mid-rapidity value of 2.9.
The momentum resolution of the spectrometer was $\Delta p/p=0.005$ at
$p=1.5$ GeV/$c$, resulting in an average
precision of better than or equal to 10 MeV/$c$ at vanishing $p_T$
for all the $Q$ variables used in the correlation analysis and
defined in sections ~\ref{para2} and ~\ref{para4}: ${\sigma}(Q_{inv})=7$
MeV/$c$,
${\sigma}(Q_{TO})=10$ MeV/$c$, ${\sigma}(Q_{TS})=5$ MeV/$c$,
${\sigma}(Q_L)=3$ MeV/$c$, ${\sigma}(Q_{T})=8$ MeV/$c$,
${\sigma}(Q_{0})=5$ MeV/$c$, ${\sigma}(Q_3)=7$ MeV/$c$. 

Severe track quality cuts were applied, resulting in a final sample of
7.9$\times10^6$ $\pi^-$, providing 13.7$\times10^6$ pairs and
13.1$\times10^6$ triplets for the Bose-Einstein correlation analysis.

\section{Single particle spectra and hydrodynamical model}
\label{sec:sps}

The study of inclusive distributions of single particles produced in
heavy-ion collisions can be interpreted in the context of models of the
source using hydrodynamical expansion.
Within the context of such models, which assume local thermal
equilibration, parameters like the temperature and collective velocity at
freeze-out can be determined.
In the limit of a stationary fireball, the distribution takes the
simple form \cite{sch93_}
\begin{equation}
E \frac{dN}{dp^{3}} \equiv \frac{dN}{ m_{T} dm_{T} dy d\phi}
\propto E
\e^{-(E-\mu)/T}
\label{equ:one_particle_spectrum_thermal}
\end{equation}
where $p$ is the Cartesian particle momentum, $m_{T} \equiv \sqrt{p_{T}^{2}+m_
{0}^{2}}$ is the transverse mass, $p_{T}$ is the transverse momentum,
$m_{0}$ is the 
rest mass, $y$ is the rapidity, 
$E = m_{T}\cosh(y-y_{fireball})$, $\mu$ is the
chemical potential and $T$ is the temperature. 
In the 
limit where only a narrow rapidity interval close to $y_{fireball}
$ is measured, the spectrum becomes
\begin{displaymath}
E \frac{dN}{dp^{3}} \sim m_{T} \e^{-m_{T}/T}
\label{equ:one_particle_spectrum_thermal_narrow_rapidity}
\end{displaymath}
and in the case of a rapidity-integrated spectrum
\begin{displaymath}
\frac{dN}{m_{T} dm_{T}} \propto m_{T} K_{1} \left( \frac{m_{T}}{T} \right) \ 
\longrightarrow \ \sim \sqrt{m_{T}} \; \e^{-m_{T}/T}
\label{equ:one_particle_spectrum_thermal_rapidity_integrated}
\end{displaymath}
where $K_{1}$ is a modified Bessel function. The last approximation holds in 
the limit $m_{T} \gg T$. Plotted against $m_{T}-m_{0}$, all particles from a 
thermalized emitter should show the same universal exponential behaviour. 
However, different additional features like transverse hydrodynamic expansion 
or particles originating 
from the decay of resonances will distort the shape of the spectrum. 
It is noted in \cite{hei95a} that for the popular fit with a simple exponential
in $m_{T}$ (without the $m_{T}$-prefactor)
\begin{equation}
E \frac{dN}{dp^{3}} = C \e^{-m_{T}/T}
\label{equ:one_particle_spectrum_exponential}
\end{equation}
an interpretation of the resulting slope parameter in terms of a temperature 
is not possible. However, since it is found to fit the measured spectrum 
better than the previous expressions, it is useful for obtaining an
estimate of the 
inverse slope parameter $T$.

\subsection{Data analysis}
\label{ssec:sps_data_analysis}

A detailed description of the analysis of the single particle spectra
presented here can be found
in \cite{vor01_}. The correction for detector acceptance and efficiency 
applied to the measured spectrum is based on a precise simulation of the 
detector, tuned in order to reproduce the measured detector response as
accurately as possible using the VENUS \cite{venus} event generator as
input. 
Multiple scattering, decays and all other reactions within the 
detector material are taken into account. Efficiency maps depending on the 
hit position on each chamber, the particle momentum and its identity are 
applied, together with position resolution and noise simulation. 
Simulated events are then reconstructed using the same code as for real
data, 
ensuring that any software-induced systematics are also taken into
account. 
The output is then matched to the VENUS input.
This correction procedure ensures that the final result is little
sensitive to remaining contamination, such as pions in a kaon sample.
A detailed study of the systematic uncertainty has been performed and it 
is found to be small, especially as regards the slope of the spectrum
($\sim$3\%). 
The absolute normalisation on the other hand is more sensitive to detector
instabilities 
which could not be perfectly simulated and is found 
to have an estimated uncertainty of at most 20\%.

\begin{figure}[hbtp]
 \begin{center}
  \rotatebox{0}{%
   \resizebox{0.5\textwidth}{!}{%
\includegraphics*{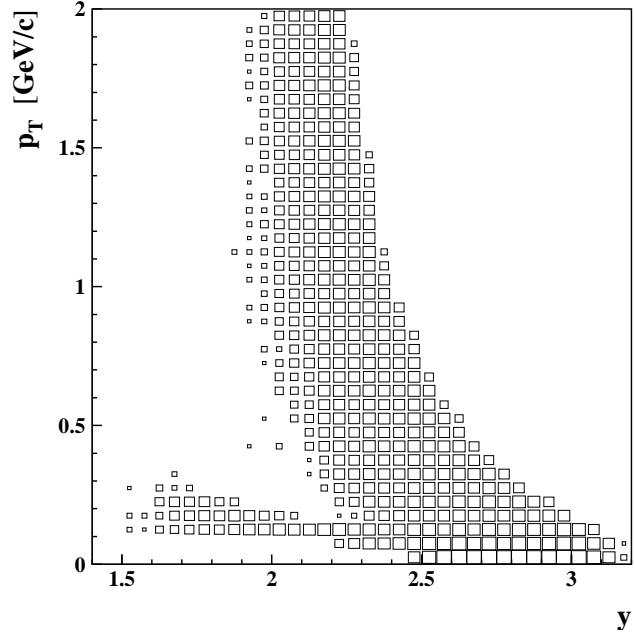}
   }
  }
  \caption{Acceptance of tracking arm~I in the $(p_{T},y)$ plane for 
pions.}
  \label{fig:2d_acceptance_pions}
 \end{center}
\end{figure}

\begin{figure}[hbtp]
 \begin{center}
  \rotatebox{0}{%
   \resizebox{0.5\textwidth}{!}{%
    \includegraphics*{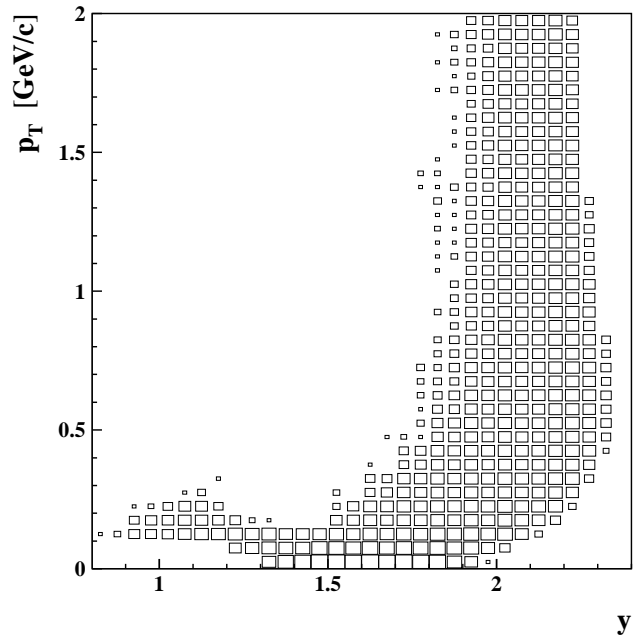}
   }
  }
  \caption{Acceptance of tracking arm~I in the $(p_{T},y)$ plane for 
kaons.}
  \label{fig:2d_acceptance_kaons}
 \end{center}
\end{figure}

Since the statistical uncertainty on the final $m_{T}$ spectrum is negligible 
compared to the systematics, it is possible to apply very severe quality cuts 
on the reconstructed tracks and on the events used. 
In contrast to the Bose-Einstein analysis, only a subset of all data is
kept using only run periods 
where the detector operation remained particularly stable. 
The analysis is performed separately for identified $\pi^{-}$ and $K^{-}$. 
The final data sample consists of 4.7$\times10^{5}$ 
$\pi^{-}$ tracks and 1.8$\times10^{4}$ $K^{-}$ drawn from
3.8$\times10^{5}$ central events. 
Figures \ref{fig:2d_acceptance_pions} 
and \ref{fig:2d_acceptance_kaons} show the detector acceptance for
negative pions and kaons respectively, obtained from the simulation.
The $\pi^-/\pi^+$ ratio and the $K^-/K^+$ ratio of the number of detected
pions and kaons have also been measured using two opposite magnetic field
polarities.
The results will be presented in a separate publication.

\subsection{Results}
\label{ssec:sps_results}

\begin{figure}[hbtp]
 \begin{center}
  \rotatebox{0}{%
   \resizebox{0.5\textwidth}{!}{%
    \includegraphics*{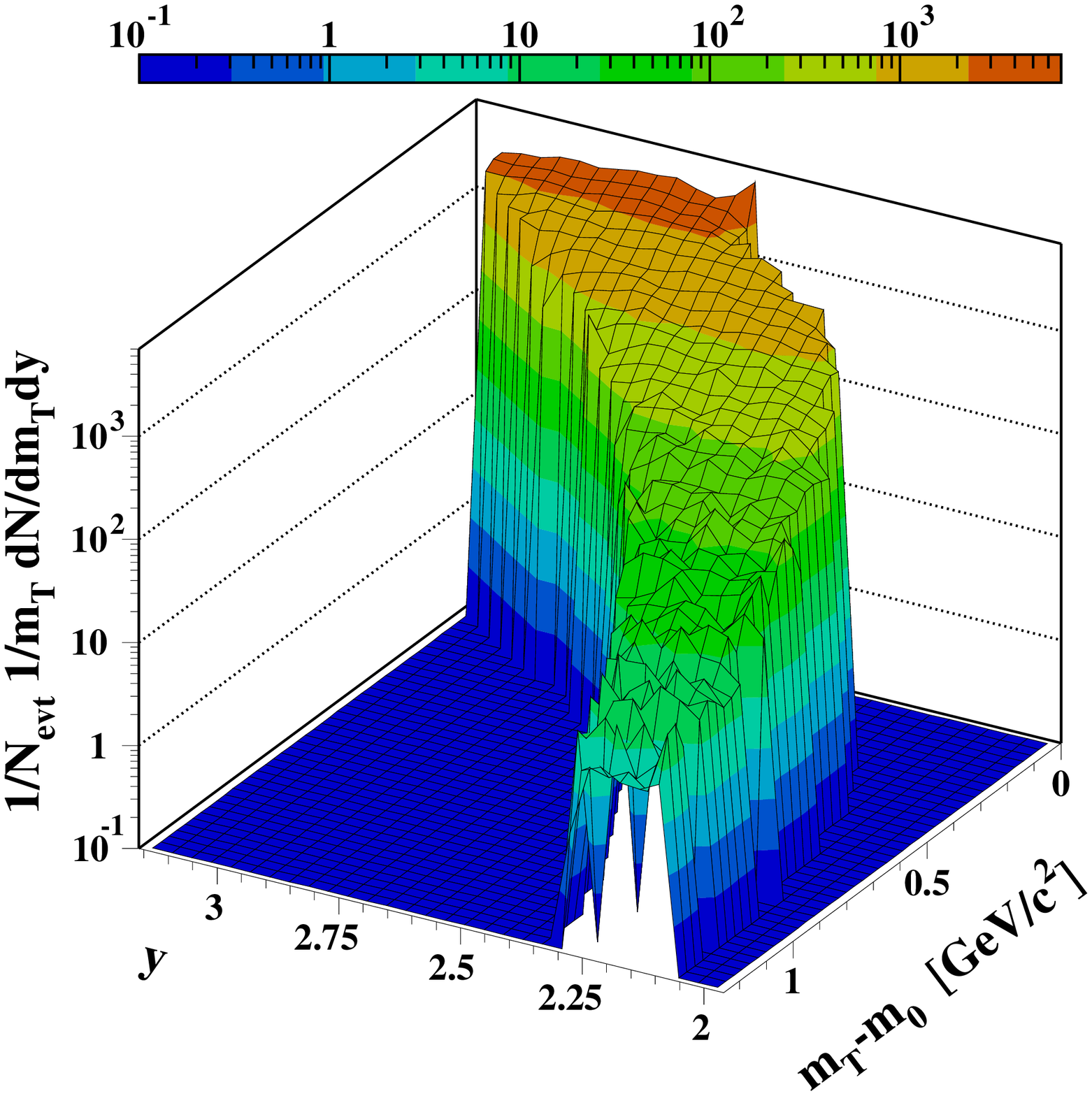}
   }
  }
  \caption{Two-dimensional (fully corrected) single pion spectrum as a 
function of the two variables $m_{T}$ and $y$.}
  \label{fig:mt_spec_2d}
 \end{center}
\end{figure}

\begin{figure}[hbtp]
 \begin{center}
  \rotatebox{0}{%
   \resizebox{0.5\textwidth}{!}{%
    \includegraphics*{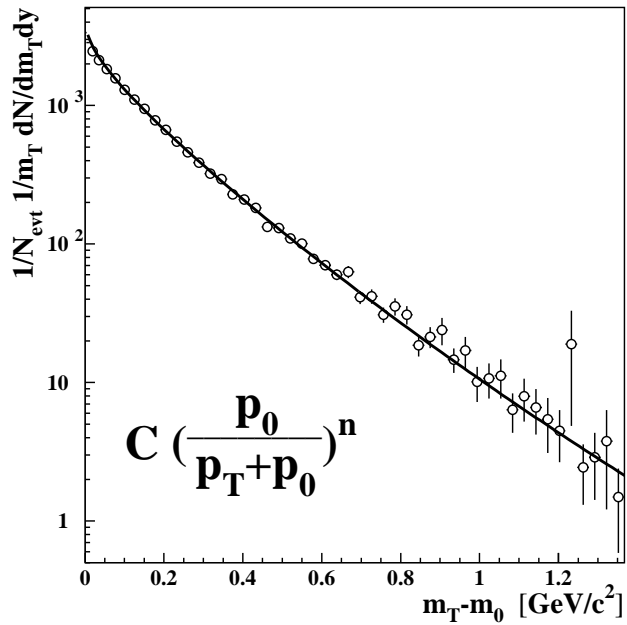}
   }
  }
  \caption{One-dimensional (fully corrected) $m_{T}$-spectrum for
$\pi^{-}$. 
The errors are statistical. The power-law fit of eq. \ref{equ:power_law_fit} 
is superimposed on the data points. The fit interval 
is $m_{T}-m_0 = [0.05,1.2] {\gevcsq}$.}
  \label{fig:mt_spec_1d_power_law_fit}
 \end{center}
\end{figure}

Figure \ref{fig:mt_spec_2d} shows the final (fully corrected) single particle 
spectrum for $\pi^{-}$ plotted as a function of the two variables $m_{T}$ 
and $y$. The form of the detector acceptance shown in 
Fig. \ref{fig:2d_acceptance_pions} is clearly noticeable, except for the
low $p_T$-low $y$ lobe, which has been omitted here.
Figure \ref{fig:mt_spec_1d_power_law_fit} shows the projection on the $m_{T}$ 
axis (normalized to a unit rapidity interval using the VENUS profile) of
the two-dimensional spectrum 
of Fig. \ref{fig:mt_spec_2d}. A fit to the exponential form of 
eq. \ref{equ:one_particle_spectrum_exponential} over the interval 
$m_{T}-m_0 = [0.1,1.2] {\gevcsq}$ (not shown in 
Fig. \ref{fig:mt_spec_1d_power_law_fit}) 
yields $C = 5220 \pm 80 \ (stat.) \ {\pma{1270}{200}} (syst.)$, and 
$T = 0.168 \pm 0.001 \ (stat.) \ {\pma{0.001}{0.005}} (syst.) {\gev}$, 
with a $\chi^{2}$ per degree of freedom of 1.6.
This result is in agreement with the NA49 result \cite{NA49SINGLE}.

At high $p_{T}$, where perturbative {QCD} becomes applicable, the spectra are 
expected to attain a power-law behaviour as observed in numerous high energy 
pp measurements (see for example \cite{boc96_}). The heavy-ion data of
this experiment seem to follow that trend even into the lower $p_{T}$ 
range. 
Therefore, a parametrization originally inspired by {QCD} \cite{hag83_} 
and successfully applied already to pp data \cite{boc96_} and heavy-ion 
data \cite{alb98_} can be used to fit the spectrum:
\begin{equation}
E \frac{d^{3}\sigma}{dp^{3}} = C \left( \frac{p_{0}}{p_{T}+p_{0}} \right)^{n}
\label{equ:power_law_fit}
\end{equation}
with $C$, $p_{0}$, and $n$ taken as free parameters. A link to the more 
familiar exponential slope parameter $T$ is obtained from the derivative of 
this expression according to
\begin{displaymath}
T_{\mrm{power-law}} = - \frac{f(p_{T})}{\frac{\partial f(p_{T})}{\partial p_{T}
}} = \frac{p_{0}}{n} + \frac{p_{T}}{n} \ .
\end{displaymath}
Thus, $p_{0}/n$ characterizes the slope of the transverse momentum spectrum 
in the limit $p_{T} \rightarrow 0$, while $1/n$ characterizes its gradient 
along $p_{T}$, i.e. the strength of the concave curvature. 
The extracted parameters are $C = 4150 \pm 70
$, $p_{0} = 4.80 \pm 1.04 {\gevc}$
and $n = 29.0 \pm 5.9$, which gives a slope parameter
$T \equiv p_{0}/n=0.166 \pm 0.005 {\gev}$. The $\chi^{2}$ per degree of
freedom 
is 1.0. The same fit performed on the $\pi^{0}$ spectrum measured by the WA98 
experiment \cite{agg00_,PIZERO,CENTRALPI,FREEZEPI} yields $C = 5120 \pm
140$, 
$p_{0} = 5.08 \pm 0.18 {\gevc}$, $n = 29.3 \pm 0.8$, and so $T = 0.173
\pm 0.002 {\gev}$.
 The acceptance for the $\pi^{0}$ measurement being different, it is 
interesting to note that fitting only the $p_{T} = [0.3,1.4] {\gevc}$ common 
interval of both spectra yields $T = 0.169 {\gev}$ for the $\pi^{-}$ 
and $T = 0.166 {\gev}$ for the $\pi^{0}$, showing that the result is stable 
with respect to the fit interval and that the spectra agree quite well.

It has also been shown that fluctuation of the parameter in an exponential
distribution leads to a final distribution of the power-like form
\cite{WILK}.
This behaviour can be interpreted in terms of a suitable
application of the nonextensive statistics of Tsallis \cite{TSALLIS}.
This interpretation is convenient to describe particle production at
fluctuating $T$, as may occur near the phase transition.
These fluctuations would exist in small parts of the hadronic system with
respect to the whole system rather than between events.
The average $\left<T\right>$ around which the temperature
fluctuates is given by $p_0/n=0.166$ GeV and the relative variance of
$1/T$ is
\begin{displaymath}
\omega = \frac{\left< (\frac{1}{T})^2\right> - \left<
\frac{1}{T}\right>^2}
{\left< \frac{1}{T}\right>^2} = \frac{1}{n} = 0.034 \ ,
\end{displaymath}
both for $\pi^{-}$ and $\pi^{0}$.
This corresponds to a nonextensivity parameter $q=1+\omega$ of 1.034.
This result, interpreted in the spirit of \cite{WILK}, indicates a
relative fluctuation $\Delta T/T$ of $18.4 \pm 1.9 \%$ ($18.4 \pm 0.3
\%$) for the $\pi^{-}$ ($\pi^{0}$) measurement.
However, this analysis neglects the contributions of resonance decays and
of flow velocity distributions.
Both of these effects increase the curvature of the pion spectrum.
Thus the value of $\Delta T/T=18.4\%$ should be considered an upper
limit on the temperature fluctuations.

The averaged negative pion yield per unit rapidity in the acceptance window is 
$1/N_{\mrm{evt}} dN/dy = 129 \pm 1 \ (stat.) \ {\pma{23}{5}} (syst.)$.
\begin{figure}[hbtp]
 \begin{center}
  \rotatebox{0}{%
   \resizebox{0.5\textwidth}{!}{%
    \includegraphics*{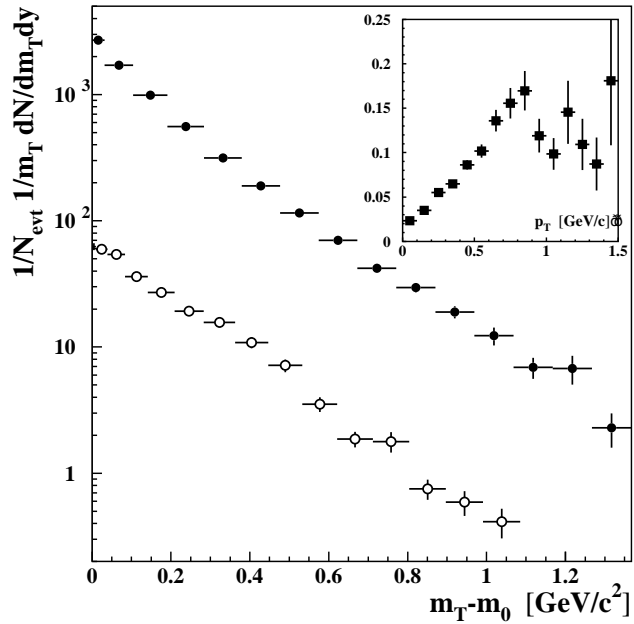}
   }
  }
 \caption{$\pi^{-}$ (black circles) and $K^{-}$ (empty circles) single 
particle spectra. The error bars are statistical only. Note that for the same 
transverse mass, the $p_{T}$ is different for pions and kaons. 
The inset shows the (bin by bin) ratio of the two $p_T$ spectra.}
  \label{fig:kaon_pion_mt_spec_comparison}
 \end{center}
\end{figure}
Figure \ref{fig:kaon_pion_mt_spec_comparison} shows the same plot as above 
for kaons. 
A fit to the exponential form \ref{equ:one_particle_spectrum_exponential} 
yields $C = 1000 \pm 120 \ (stat.) \ {\pma{260}{1}} (syst.)$, and $T = 0.181 
\pm 0.005 \ (stat.) \ {\pma{0.001}{0.009}} (syst.) {\gev}$. So the inverse 
slope $T$ for kaons and pions are comparable.
It should be noted 
that the rapidity acceptance is quite different for the two particle species, 
the $\pi^{-}$ acceptance being much closer to mid-rapidity. 
An estimate of the effect induced by this difference in acceptance can be 
made using the term $\cosh(\Delta y)$ given below eq. 
\ref{equ:one_particle_spectrum_thermal}. The inverse slope for $K^{-}$ 
would become of the order of $0.230 {\gev}$.
The $K^{-}/\pi^{-}$ ratio at a common rapidity of 2.3 is $11.4 \pm 0.4 \
(stat.) \ {\pma{0.6}{2.7}} (syst.) \%$,
which is in agreement with the NA49 result \cite{NA49SINGLE}.

\subsection{Hydrodynamical Source Expansion Model}
\label{ssec:sps_hydro_model}

The single pion spectrum has been fitted in the 
interval $0.05 < m_{T}-m_0 < 1.2 {\gevcsq}$ with the Wiedemann-Heinz
(W.-H.) 
model
\cite{cha95b,cha95c,hei96b,wie96c,wie97b,wie99a}. 
It relies on the following idea: the main characteristics of the particle 
phase-space distribution at freeze-out can be quantified by its widths in the 
spatial and temporal directions, a collective dynamical component 
(parameterized by a flow field) which determines the strength of the 
position-momentum correlations in the source, and a second, random dynamical 
component in momentum space (parameterized by a temperature). The model 
emission function contains seven parameters, but the shape of the single 
particle transverse mass spectrum is fully determined by the temperature $T$ 
and the transverse flow rapidity profile $\eta_{T}(r) = \eta_{f}
\frac{r}{R}$ which is 
assumed to depend linearly on the transverse coordinate $r$, where $\eta_{f}$ 
is the transverse flow rapidity strength, and $R$ the Gaussian transverse
spatial width.
The mean transverse flow velocity $\left<\beta_T\right>$ can be easily
calculated as the mean value of $\tanh(\eta_T(r))$ over the transverse
source profile.
The results will be given as a function of $\left<\beta_T\right>$ 
rather than $\eta_{f}$, since its physical interpretation is more
straightforward.

\begin{figure}[hbtp]
 \begin{center}
  \rotatebox{0}{%
   \resizebox{0.5\textwidth}{!}{%
\includegraphics*
{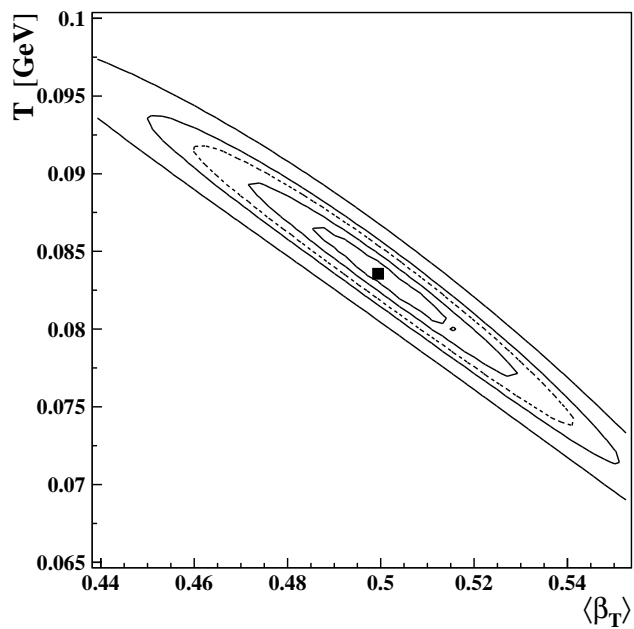}
   }
  }
  \caption{$\chi^{2}$ contour plot of the hydrodynamical model fit to the 
measured $\pi^{-}$ single particle spectrum of Fig. 
\ref{fig:mt_spec_1d_power_law_fit}. A Gaussian profile is assumed for the 
transverse spatial profile of the source. 
The curves displayed represent successively (starting with the innermost one) 
contours at 50\% CL, 90\% CL, 99\% CL (dashed line), 99.9\% CL, and the last 
curve indicates a highly excluded region. The black square represents
the best 
parameter values.}
  \label{fig:plot_temp_etaf_gauss_nores}
 \end{center}
\end{figure}

\begin{figure}[hbtp]
 \begin{center}
  \rotatebox{0}{%
   \resizebox{0.5\textwidth}{!}{%
\includegraphics*
{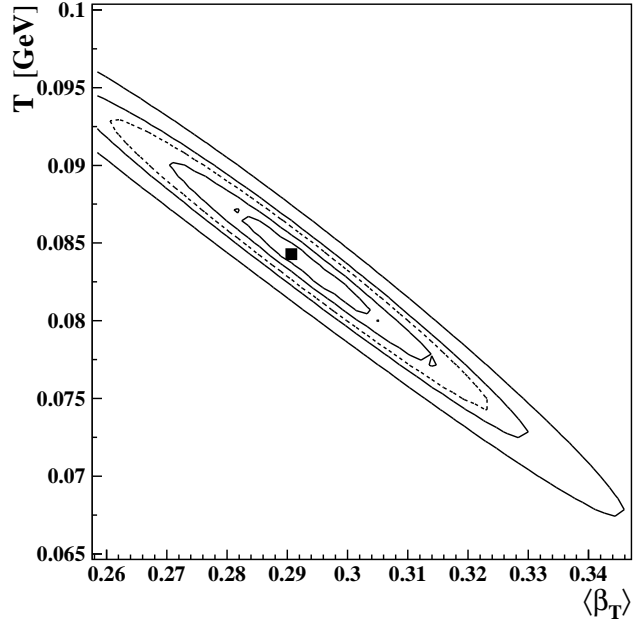}
   }
  }
  \caption{Same plot as in Fig. \ref{fig:plot_temp_etaf_gauss_nores}, but 
assuming a box profile for the transverse spatial profile of 
the source.}
  \label{fig:plot_temp_etaf_box_nores}
 \end{center}
\end{figure}

The three parameters that are allowed to vary freely during the fitting 
procedure are $T$, $\eta_{f}$, and an overall normalization factor.
The result of the fit is $T = 0.084 \pm 0.003 {\gev}$ and
$\left<\beta_T\right> =
0.50 \pm 0.02$, the $\chi^2/d.o.f.$ being 1.1. 
The resulting curve is not shown in Fig.
\ref{fig:mt_spec_1d_power_law_fit} 
as it is hardly distinguishable from the power-law fit. 
The temperature and flow parameters are strongly correlated, 
so it is more interesting to consider a $\chi^{2}$ contour plot of the fit to 
the measured single particle spectrum as a function of those two parameters. 
Figure \ref{fig:plot_temp_etaf_gauss_nores} shows the result assuming a 
Gaussian transverse spatial profile, and Fig. 
\ref{fig:plot_temp_etaf_box_nores} assuming a box profile of the same rms
width as the Gaussian one. 
Only direct pions are considered in the model. 
Furthermore, only statistical errors are taken into account during the
fit.
The curves displayed represent 
different confidence levels for the $(T,\left<\beta_T\right>)$ values. 
The dependence of the $\chi^2/d.o.f.$ on the shape of the source is very
small, the best
$\chi^2/d.o.f.$  
being slightly larger for the box profile (1.5).
While the best fit temperature appears to be independent of the shape of
the source, the best fit flow velocity is considerably smaller
($\left<\beta_T\right>=0.29$) for the box profile.
Inclusion of the resonance decay contribution in the model calculation
would be expected to reduce the extracted flow velocity parameters and
increase the temperature parameters.
The precise effects do however depend on details of the models used.

\section{Two-pion correlations}
\label{para2}

Bose-Einstein interferometry is most commonly used to study pairs of
identical particles.
The correlation function $C_2$ is defined, up to a proportionality factor
$\cal N$, as the ratio of a two-particle spectrum ${\cal P}_2$ over the
product of two single particle spectra ${\cal P}_1$:
\begin{displaymath}
C_2({\vec{p}}_{1},{\vec{p}}_{2}) = {\cal N}\;
\frac{ {\cal P}_2({\vec{p}}_{1},{\vec{p}}_{2})}
     { {\cal P}_1({\vec{p}}_{1}) \cdot{\cal P}_1({\vec{p}}_{2})}
\end{displaymath}
with 
\begin{displaymath}
{\cal P}_1({\vec{p}}) = E\;\frac{ dN}{d{p}^3}
\end{displaymath}
and 
\begin{displaymath}
{\cal P}_2({\vec{p}}_{1},{\vec{p}}_{2}) = E_1\;E_2\;\frac{
dN}{dp^{3}_{1}\;dp^{3}_{2}}
\end{displaymath}
where $E_i$ and $\vec{p}_{i}$ are the energy and momentum of
particle $i$, respectively.

Experimentally, the product of one-particle distributions in the
denominator is commonly obtained by a mixed event technique  whereas the
two-particle distribution in
the numerator is constructed from all pair combinations of
identical particles found in each event.
$C_2$ is then normalized to unity far away from the interference region. 

A fully chaotic source can be seen as a superposition of uncorrelated
elementary sources, and one- and two-particle distributions may be
expressed through the Wigner function of the source $S(x,k_{12})$
\cite{ZAK}.
The correlation function is then written
\begin{displaymath}
C_2({\vec{p}}_{1},{\vec{p}}_{2})  =
1\;+\;\lambda\;\frac{|\int\!\:d^4x\;S(x,k_{12})\;\exp[iq_{12}x]|^2}{|\int\!
\:d^4x\; S(x,k_{12})|^2}
\label{eq:fourier}
\end{displaymath}
with $q_{12}=p_{1}-p_{2}$, the 4--momentum difference of the two
particles, $k_{12}=(p_{1}+p_{2})/2$ and $x=(x_{1}+x_{2})/2$, the mean
space-time coordinate of the pair emission point.
The chaoticity parameter $\lambda$ is inserted 
to take into account the possibility that the source may
not be fully chaotic and also that any wrongly reconstructed tracks, or
tracks originating from decays of long-lived resonances, will
dilute the Bose-Einstein correlations in the data.

A one-dimensional interferometry analysis is commonly made as a function
of $Q_{inv}\equiv\sqrt{-q_{12}^2}$, whereas a multidimensional analysis
uses a set of $Q$ variables which are defined as various projections of
$q_{12}$.

\subsection{Data analysis and results}

Two independent analyses were performed with the complete data set
recorded in
the negative tracking arm \cite{vor01_,QM01}
totalling 13.7$\times10^6$ pairs of identified $\pi^-$.
These data were corrected for resolution and Coulomb effects in an
iterative way \cite{PRATT}.
The Gamow correction was not used as it overcorrects the data in the
$Q_{inv}$ range of 0.1 to 0.3 GeV/$c$.
Because the finite resolution in the measurement of the $Q$ variables
leads to
an underestimate of the radii and $\lambda$ parameters, a correction has
to be implemented in the fitting procedure.
This is done by a convolution method, replacing the $C_2(\vec{Q})$ formula
expressing the two-particle correlation function
used to fit the data by 
$$C_2^{rc}(\vec{Q})=\int\!\!\int\!\!\int r(\vec{Q},\vec{Q'})\:
C_2(\vec{Q'})\:d\vec{Q'}$$
where $r(\vec{Q},\vec{Q'})$ is the resolution function which is chosen to
be Gaussian:
$$\hspace{-3.0cm}r(\vec{Q},\vec{Q'})=1/(2\pi)^{3/2}\:1/|V|^{1/2}\:$$
$$\hspace{1.5cm}\times\exp[-1/2\:(\vec{Q}-\vec{Q'})^T\:V^{-1}\:(\vec{Q}-\vec{Q'})]$$

\begin{figure}[hbtp]
\begin{center}
\hspace{-0.35cm}
\resizebox{0.5\textwidth}{!}{%
  \includegraphics{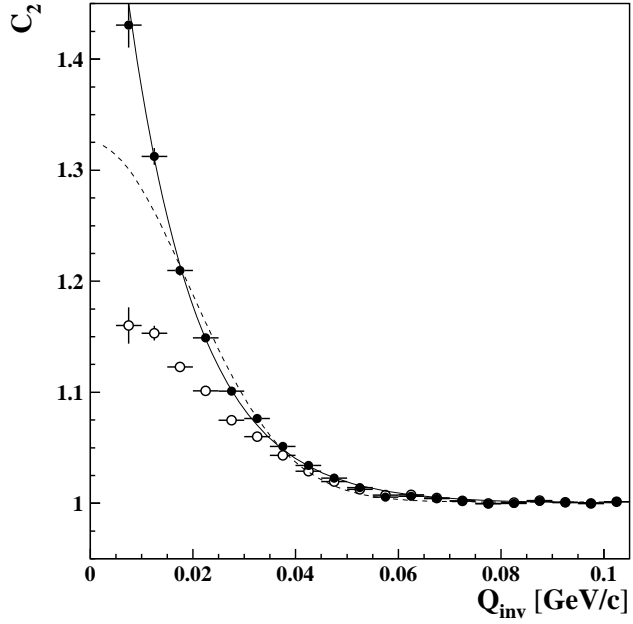}
}
\caption{The measured two-pion correlation function $C_2$ (full symbols),
 corrected for resolution and Coulomb
effects, as a function of
$Q_{inv}$. The full line is a fit to an exponential form 
whereas the dashed curve is a fit to a Gaussian form.
The empty symbols show the data before corrections.
} 
\label{fig:figure6}
\end{center}
\end{figure}

The diagonal elements of the covariance matrix $V$ are set equal to the
square of the various $Q$-resolutions, which are estimated by a full
simulation of the experimental setup as a function of the $k_T=|\vec
p_{T1}+\vec
p_{T2}|/2$ of the pairs.
The non-diagonal elements of $V$ are neglected, as the resolution
correction has a very small effect compared to the Coulomb
correction.
Figure ~\ref{fig:figure6} shows $C_2$, the measured $\pi^-$$\pi^-$ correlation
function,
plotted as a function of $Q_{inv}$
before and after the resolution and Coulomb corrections.
In the plot, the correction for resolution is obtained by multiplying each data
point by
$C_2^{rc}(Q_{inv})/C_2(Q_{inv})$.
$C_2$ is clearly exponential
\cite{BEWA98}.
The solid curve is a fit of the form 1+$\lambda_e$exp[-2$Q_{inv}R_{inv}]$ 
which gives
$R_{inv}=7.33\pm0.08$ fm and $\lambda_e=0.788\pm0.009$ for
$\left<k_T\right>=0.116$
GeV/$c$.
This exponential behaviour appears not to hold in the 3-d analysis, where
the projected slices are better represented by Gaussians.
\begin{figure}[hbtp]
\begin{center}
\hspace{-0.7cm}
\resizebox{0.50\textwidth}{!}{%
  \includegraphics{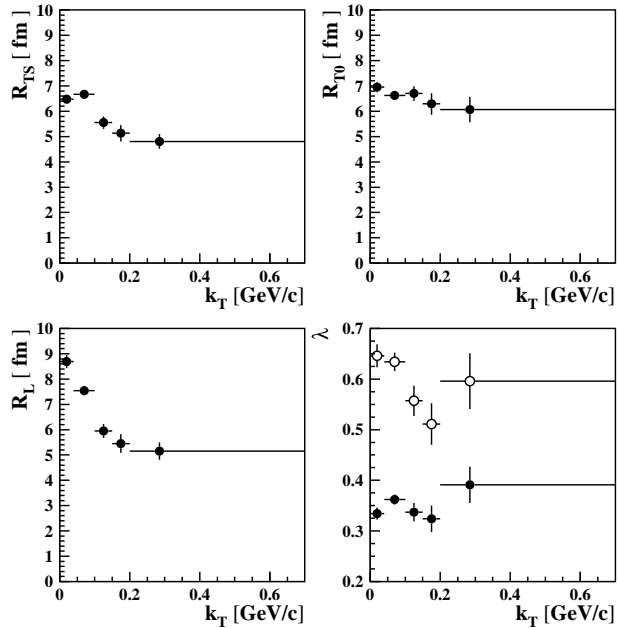}
}
\vspace*{-0.4cm}
\caption{The $k_T$ dependence of the Pratt-Bertsch parameters.
The points are plotted at the average $k_T$ of the bin and the horizontal 
bars indicate the bin width. The open symbols in the figure at bottom
right
show the $\lambda$ parameter after correction for background from
misidentified pions, as explained in section ~\ref{para}.} 
\label{fig:figure7}
\end{center}
\end{figure}
The 3-d analysis of Bose-Einstein correlations has been done
using two different parameterizations in the Longitudinally CoMoving
System
(LCMS): the standard Pratt-Bertsch fit (PB) in the 3-dimensional
space of momentum
differences $Q_{TS}$ (perpendicular to the beam axis and to
the transverse momentum of the pair), $Q_{TO}$ (perpendicular
to the beam axis and parallel to the transverse momentum of
the pair), and $Q_{L}$ (parallel to the beam axis) \cite{stand}, including
a cross term
$R^2_{out-long}$ as predicted \cite{PB}
$$C_2=1+\lambda\exp[-Q_{TS}^2R_{TS}^2-Q_{TO}^2R_{TO}^2-Q_L^2R_L^2$$
$$\hspace{3.9cm}-2Q_{TO}Q_LR^2_{out-long}]$$
and the Yano-Koonin-Podgoretski{\u{\i}} fit (YKP) \cite{cha95b} in the
$Q_0$ 
(energy difference of the pair), $Q_T$,$Q_L$ space
$$C_2=1+\lambda\exp[-Q^2_TR^2_T+(Q^2_0-Q^2_L)R^2_4-(Q{\cdot}U)^2(R^2_0
+R^2_4)]$$
where $U=\gamma(1,0,0,v_L)$, $\gamma=1/\sqrt{1-v_L^2}$ with $v_L$ in units
of $c=1$.
In the YKP parameterization the different radii are invariant under a
longitudinal Lorentz boost, and the speed parameter $v_L$ connects the
local rest frame to the measurement frame (the LCMS in our case).
The results as a function of the $k_T$ of  the
pairs are shown in Figs.~\ref{fig:figure7} and ~\ref{fig:figure8} and
summarized in Table~\ref{tab:table1}.
The $\lambda$ parameters of the YKP fit (not shown in
Table~\ref{tab:table1})
are found compatible with the $\lambda$ parameters of the PB fit.
\begin{figure}[hbtp]
\begin{center}
\hspace{-0.7cm}
\resizebox{0.50\textwidth}{!}{%
  \includegraphics{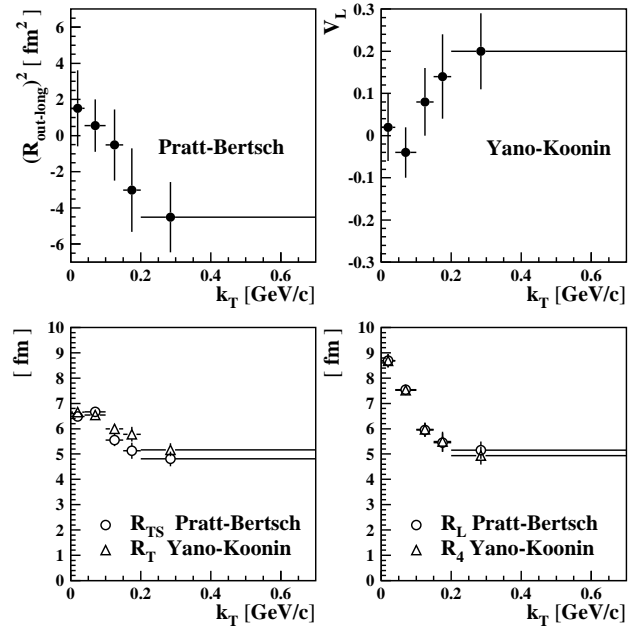}
}
\vspace*{-0.4cm}
\caption{Comparison between the PB and YKP fits in the LCMS.
The cross term $R^2_{out-long}$ from the PB fit and $v_L$ from the YKP
fit both show a deviation from zero at large $k_T$ when estimated in the 
LCMS.
The $R_{TS}$ and $R_{L}$ parameters from the PB fit are in good
agreement with respectively $R_T$ and $R_4$ from the YKP fit.
} 

\label{fig:figure8}
\end{center}
\end{figure}

The systematic errors, not included in
Figs.~\ref{fig:figure7} and ~\ref{fig:figure8}, and
Table~\ref{tab:table1}, are estimated by varying the
different analysis cuts, including the cuts used to identify the pion with
the time of flight system.
The systematic error on the Coulomb correction due to the error on the
determination of the radius parameters is also taken into account.
All these variations are added in quadrature.
The total relative systematic errors on the radii $\Delta{R}/R$ amount to
0.8\% for $R_{inv}$, 1.4\% for
$R_{TS}$, 3.5\% for $R_{TO}$, 9.1\% for $R_{L}$, 0.8\% for $R_{T}$, and
9.7\% for $R_{4}$.
Systematic errors on $R^2_{out-long}$ and $v_L$ are asymmetric and reach
respectively $\pma{2.2}{1.7}$ fm$^2$ and $\pma{0.13}{0.08}$.

The $R_{TS}$ and $R_{L}$
parameters from the PB fit are in good agreement
respectively with $R_T$ and $R_4$ from the YKP fit. The cross term
$R^2_{out-long}$ from the PB fit and $v_L$ from the YKP
fit deviate from zero. In a source undergoing a boost invariant
expansion
the local rest frame coincides with the LCMS. Both the
cross term and $v_L$
expressed in the LCMS are then expected to vanish \cite{cha95b}. As this 
is not quite the case, it suggests that the source seen within the
acceptance does not have a strictly boost invariant expansion.
The strong decrease of the longitudinal radius $R_L$ or $R_4$ with
$k_T$ compared to the transverse radii $R_T$,
$R_{TS}$, $R_{TO}$ shows a longitudinal expansion which is larger than
the transverse one.
Finally, the $R_0$ parameter (not shown in the figures), which corresponds
to the duration
of emission of particles from the source, is compatible with zero for all
$k_T$ bins, excluding a long-lived intermediate phase.
These results agree with the previously published WA98 results obtained
using roughly half of the present data sample   
\cite{BEWA98}.
The WA98 analysis is compatible with the NA49 results obtained
in a
slightly different rapidity range ($\left<y\right>=3.2$) with
unidentified
negative particles \cite{NA49}.

\subsection{Comparison with a hydrodynamical model based simulation}

\begin{figure}[hbtp]
\begin{center}
\hspace{-0.7cm}
\resizebox{0.50\textwidth}{!}{%
  \includegraphics{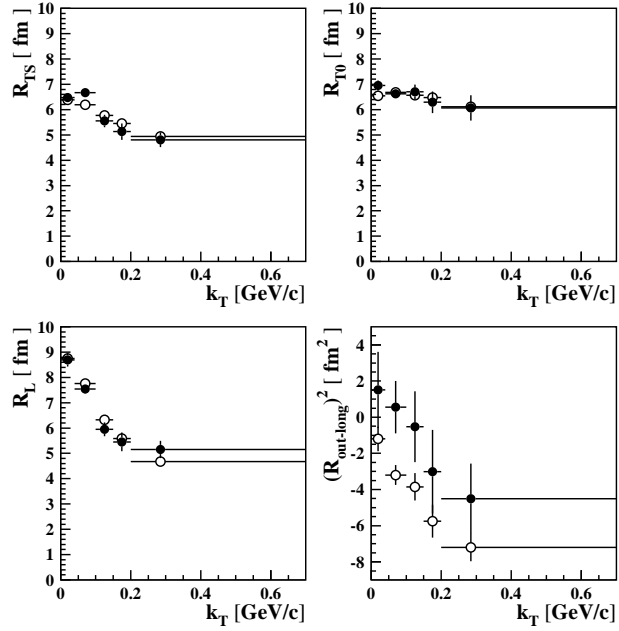}
}
\vspace*{-0.4cm}
\caption{Comparison between simulation and data as a function of $k_T$. 
The open symbols
are the result of the fit of the PB formula to the correlation functions
produced by the
hydrodynamical model, whereas the full symbols are the result of the fit
to the data. }

\label{fig:figure14}
\end{center}
\end{figure}

The W.-H. model can be used to generate correlation functions which
can be compared to the data. 
For that purpose, all necessary integrations are performed numerically
\cite{wie97b} to get the value of $C_2$ in the PB parameterization for
given values of $Q_{TS}$, $Q_{TO}$, $Q_L$, $k_T$, and the rapidity
$Y=(y_1+y_2)/2$ of a pair.
To simulate properly the acceptance of the negative tracking arm, the mean
values of $k_{T}$ and $Y$ for each $(Q_{TS},Q_{TO},Q_L)$ bin are calculated
using real data, and then for each bin the mean values 
$(\left< Q_{TS} \right> ,\left< Q_{TO} \right> ,
\left< Q_L \right>, \left< k_{T} \right> , \left<Y \right>)$ 
are used as input for the hydrodynamical calculation.
This procedure is repeated for all five $k_{T}$ intervals used in the data
analysis.
The correlation functions are generated neglecting contribution from
resonances,
using $T = 85$ MeV, $\eta_{f} = 0.5$
 (values extracted from the single particle spectra analysis), $R = 8$ fm,
$\tau_{0} = 11$ fm/$c$, $\Delta\tau = 2$ fm/$c$ and $\Delta\eta =
1.3$ (see \cite{wie97b} for definitions of the parameters).
Figure~\ref{fig:figure14} shows the Bose-Einstein radii
extracted from fitting the simulated correlation functions with the PB 
formula.
The error bars used to perform these fits are taken
to be the same as the ones calculated for the real data in each
$(Q_{TS},Q_{TO},Q_L)$ bin.
The measured results are shown on the same plot, and the agreement is
found to be good.
The shift in the the cross term $R^2_{out-long}$ between data and
simulation is compatible with the systematic uncertainty.

\section{Average pion phase-space density at freeze-out}
\label{para}

As the $m_T$ spectrum gives the momentum-space density at freeze-out and
as the Bose-Einstein correlation radii provide information on the
covariant volume for particles of momentum $\vec{p}$, it is possible, by
combining these results, to extract the average phase-space density
$\left<f\right>(p_T,y)$ at freeze-out \cite{BERTSCH,BARRETTE}:
\begin{displaymath}
\left<f\right>=
\frac{\sqrt{\lambda}}
{(\frac{E_\pi}{\pi^{3/2}})R_{TS}\sqrt
{R_{TO}^2R_L^2-R^4_{out-long}
}} {\ }
\frac{dn}{{dy}{\ }{p_T}{\ }{dp_T}{\
}{d\phi}}
\end{displaymath}
with $E_{\pi}=\sqrt{m_{\pi}^2+p^2}=m_T\cosh{y}$.
The radii and $\lambda$ are functions of $p_T$ and $y$.
The factor $\lambda$, which comes from the two-pion correlation analysis, 
corrects for contributions of pions originating from
long-lived resonances decaying close to the primary vertex.
A difficulty of this method is to include only the contribution
of the real pions in the determination of $\lambda$, excluding backgrounds
from misidentified particles.
This is achieved by applying to $\lambda$, separately for each $k_T$ bin, 
a correction
factor obtained from a full simulation of the experimental setup, taking
into account geometrical acceptance, backgrounds and efficiency of the
chamber-camera-time of flight system.
The effects can be seen in Fig.~\ref{fig:figure7}, bottom
right, where $\lambda$ with and without correction is displayed.
\begin{figure}[hbtp]
\begin{center}
\hspace{-0.35cm}
\resizebox{0.5\textwidth}{!}{%
  \includegraphics{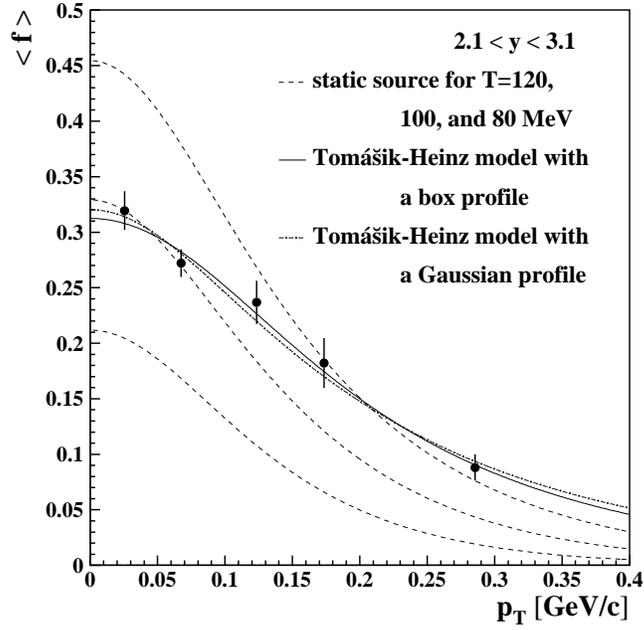}
}
\caption{Average pion phase-space density as a function of $p_T$ compared
to different models (see text).
} 
\label{fig:figure9}
\end{center}
\end{figure}
Figure~\ref{fig:figure9} and Table~\ref{tab:table2} show the results on
the average phase-space density for $\pi^-$ as a function of $p_T$.
The error bars reflect the statistical errors only.
The systematic uncertainties are dominated by the uncertainty of the
correction on the $\lambda$ parameter, which is estimated to be 20\%,
and by the systematic error on $R_L$, giving a total of 13.7\%
systematic error on $\left<f\right>$.
Within errors, all nuclear collision measurements at the SPS are found
to be indistinguishable \cite{HWC}, and our result at
mid-rapidity agrees well with these previous measurements.
The dashed lines indicate Bose-Einstein
density distributions $\left<f\right>=[\exp(E_\pi/T)-1]^{-1}$
of static sources of pions ($E_\pi\approx m_T$) for three choices of the
freeze-out temperature $T$:
80 MeV, 100 MeV, and 120 MeV, from the lower to the higher curve.
The results are in rough agreement with the 100 MeV distribution at low
$p_T$,
but show a clear deviation from a Bose-Einstein
distribution at high $p_T$.
As pointed out in \cite{T-H}, this deviation is mainly due to the strong
longitudinal expansion of the
fireball which reduces the spatially averaged phase-space density and, to
a lesser extent, due to the radial collective flow, which adds extra
transverse momentum to the particles compared to particles emitted by a
static source.
Consequently, even in the absence of transverse flow, $\left<f\right>$ 
will be reduced compared to a Bose-Einstein density distribution.
This effect may necessitate a positive pion chemical
potential in order to match the experimental observation.
Such a positive potential can be related to the presence of pions from
short-lived resonance decays.
The Tom\'a{\u{s}}ik-Heinz (T.-H.) model \cite{T-H} used to fit the data
assumes a thermalized fireball with a longitudinally boost-invariant
expansion and a transverse flow rapidity profile which depends linearly on 
the transverse coordinate.
This model includes a pion chemical potential and has three free
parameters, the freeze-out temperature $T$, the strength of the transverse
flow rapidity profile $\eta_t$, and $\mu_0$, the chemical potential value
in the center of the fireball. 
In Fig.~\ref{fig:figure9}, the full line is a fit of the T.-H. model with
a box transverse density profile, whereas the point-dashed
line is a fit of the same model with a Gaussian transverse profile.
Both fits agree well with the data with a $\chi^2/d.o.f.$ of 0.78 and 0.93
for respectively the box and the Gaussian profiles,
giving $T=87\pma{52}{27}$ MeV,
$\eta_t=0.49\pma{0.07}{0.22}$ and $\mu_0=57\pma{29}{125}$ MeV 
for the box profile.
This $\eta_t$ result corresponds to a mean transverse flow velocity 
$\left<\beta_T\right>=0.42\pma{0.05}{0.18}$.
\begin{figure}[hbtp]
\begin{center}
\hspace{-0.35cm}
\resizebox{0.5\textwidth}{!}{%
  \includegraphics{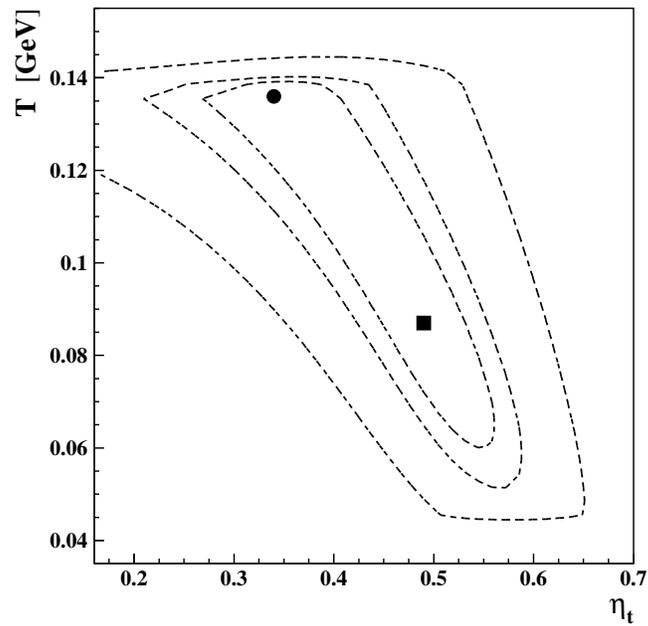}
}
\caption{$\chi^2$ contour plot for the T.-H. model with a box profile.
Starting from the centre, the curves represent contours at 39\%, 70\% and
99\% confidence level for $T$ and $\eta_t$.
The black square is the best fit ($\mu_0=57$ MeV).
The full circle is the result with $\mu_0$ excluded from the fit and set
to zero.
} 
\label{fig:figure10}
\end{center}
\end{figure}
Due to the lack of experimental points at large $p_T$, the T.-H. model
with
the Gaussian profile provides very loose estimates of these parameters,
which are nevertheless compatible with those from the box profile.
Figure~\ref{fig:figure10} shows the $\chi^2$ contour plot for the T.-H.
model with the box
profile for $T$ and $\eta_t$.
The curves represent (starting from the centre) contours at 39\%, 70\% and
99\% confidence level.
The best fit is obtained for a pion chemical potential $\mu_0=57$ MeV
(black square)
but a solution with $\mu_0=0$ MeV is also possible (full circle).
On the other hand, a large pion chemical potential at freeze-out, close to
the Bose condensation limit of $\mu_0=m_\pi$, as could be speculated
given the rather low mass of the pion, seems to be excluded in view of the 
error bar on $\mu_0$.
The results of the fit of the W.-H. model on the single $\pi^-$ spectrum
can be compared to the fit of the T.-H. model on the phase-space
distribution.
The agreement is good for $T$, and satisfactory for $\left<\beta_T\right>$ 
when taking into account the systematic error. It should be noted that
the constraint provided by the fit of the
T.-H. model is weak compared to the one given by the W.-H. model because
of the relatively small amount of experimental points in the phase-space
distribution.
Moreover the T.-H. model uses a pion chemical potential
whereas the W.-H. model doesn't.
Finally, as $\left<f\right>$ is the pion occupation per 6-d
position$\otimes$momentum
cell, the obtained values do not provide striking evidence for
the presence of an excess of pions or for the presence of large
disoriented chiral condensates.
This chiral condensate phenomena has been investigated by
other means within WA98
\cite{DCC1,DCC2}.

\section{Three-pion correlations}
\label{para4}

In the hypothesis of a fully chaotic source of identical particles, the
two-pion correlation function can be written $C_2=1+|F_{12}|^2$
where $|F_{12}|^2$ is the Fourier transform squared of the space-time
source function.
The three-pion correlation function is then 
$C_3=1+|F_{12}|^2+|F_{23}|^2+|F_{31}|^2+2\cdot \mbox{Re}\{ F_{12}\cdot
F_{23}\cdot F_{31}\}$.
The terms $|F_{ij}|^2$, which express the contribution of the three
combinations of
the two-pion correlations contained in the triplet (123), 
provide the largest contribution to $C_3$.
The last term only represents the genuine three-body correlation.
It can be written $2\cdot |F_{12}|\cdot
|F_{23}|\cdot
|F_{31}|\cdot W$, where $F_{ij}$ is defined as
$|F_{ij}|\exp[i\phi_{ij}]$ and $W\equiv
\cos(\phi_{12}+\phi_{23}+\phi_{31})$.
The simultaneous measurement of $C_2$ and $C_3$ provides information on
$W$, the cosine of the sum of the three phases of the Fourier transforms.
In contrast, the measurement of $C_2$ alone gives access only to the
radii of the source and not to the phases.
If the emission source is fully chaotic, $W$ measures the asymmetry
of the source.
In the presence of not fully chaotic sources, which is likely to be the
case, $W$, the strength of the true three-body correlation, is basically
sensitive to the coherence.
So, in the framework of the partially coherent model \cite{HEINZ}, $W$
gives information on the degree of chaoticity of the emission source
in a manner which is insensitive to backgrounds such as the contributions
from resonances.
The complete data set recorded with the negative tracking arm yielded a
total of 13.1$\times10^6$ triplets of $\pi^-$.
\begin{figure}[hbtp]
\begin{center}
\hspace{-0.35cm}
\resizebox{0.5\textwidth}{!}{%
  \includegraphics{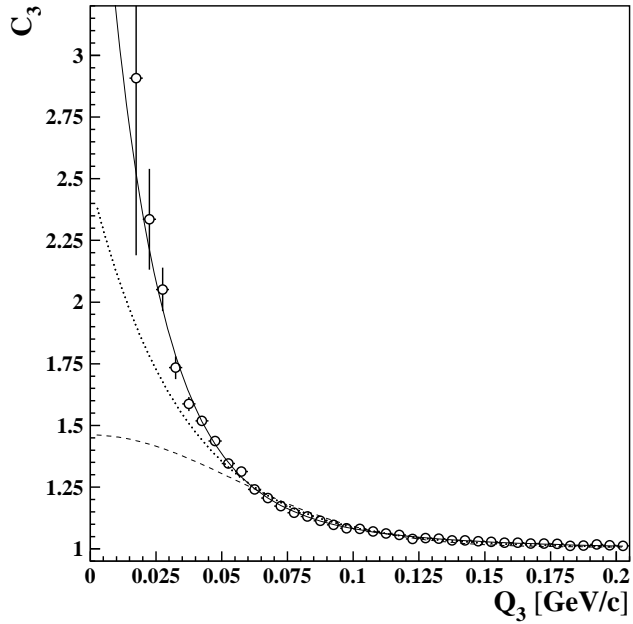}
}
\caption{The three-pion correlation function $C_3$ as a function of
$Q_3$. The full line is a fit to a double exponential form (see text).
The dotted and the dashed lines are fits to respectively a single exponential and a Gaussian form.} 
\label{fig:figure12}
\end{center}
\end{figure}
After correction for resolution and Coulomb effects
\footnote{
The Coulomb correction applied to a particular triplet is
the product of the Coulomb corrections used for the three pair
combinations contained in that triplet.},
a strong $C_3$
signal is observed (Fig.~\ref{fig:figure12}) as a function of
$Q_3\equiv\sqrt{Q_{12}^2+Q_{23}^2+Q_{31}^2}$ with
$Q_{ij}\equiv\sqrt{-(p_i-p_j)^2}$, which can be fitted by a
double exponential function
$$C_3=1+\lambda_1\exp[-2Q_3R_1]+\lambda_2\exp[-2Q_3R_2]$$
with fitted parameters $R_1=5.08\pm0.26$ fm, $\lambda_1=3.12\pm0.27$,
$R_2=1.66\pm0.08$ fm,
$\lambda_2=0.341\pm0.046$ and $\chi^2/d.o.f.=1.10$.
Such non-Gaussian behaviour was in fact predicted by a final-state
rescattering model \cite{HUMAN}.   
After the measurement of $C_2$ and $C_3$, the data are analysed again and
the experimental value of $W$
is calculated using
\begin{displaymath}
W=\frac{ \{C_3(Q_3)-1\}-\{C_2(Q_{12})-1\}-\{C_2(Q_{23})-1\}-
\{C_2(Q_{31})-1\} }
   { 2\cdot\sqrt{ \{C_2(Q_{12})-1\}\{C_2(Q_{23})-1\}\{C_2(Q_{31})-1\} } }
\end{displaymath}
individually for each triplet found, characterized by $Q_3$ and by the
values $Q_{12}$, $Q_{23}$, and $Q_{31}$ corresponding to the three pair
combinations contained in the triplet.
\vspace*{-.2cm}
\begin{figure}[hbtp]
\begin{center}
\hspace{-0.35cm}
\resizebox{0.5\textwidth}{!}{%
  \includegraphics{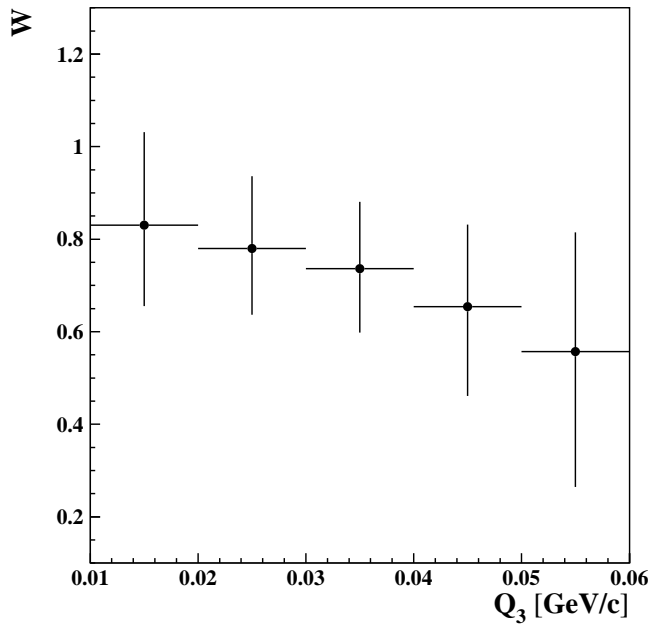}
}
\caption{The factor $W$ as a function of $Q_3$. The error bars
include statistical and systematic errors. The statistical errors alone 
are contained within the size of the symbols.
} 
\label{fig:figure13}
\end{center}
\end{figure}

As described earlier, the estimate of systematic errors is done by varying
the different analysis cuts, both in the two- and three-pion correlation
analysis.
The effects on $W$ of the statistical errors in the measurement of $C_2$
and $C_3$ are treated as systematic errors by changing $C_2$ ($C_3$) by
$\pm\sigma_{C_2}$ ($\pm\sigma_{C_3}$).
This last source of error dominates the other ones.
All these variations are then added in quadrature.

Figure~\ref{fig:figure13} and Table~\ref{tab:table3} show $W$ as a
function of $Q_3$ for $Q_3\leq60$
MeV/$c$, beyond which the $W$ significance is too low.
The error bars include statistical and systematic errors.
The statistical errors (not shown separately in Fig.~\ref{fig:figure13})
are by comparison negligible.
In view of the errors, no significant $Q_3$ dependence is observed.
The genuine three-pion correlation is found to be substantial with a
weighted mean     
over the five bins $\left<W\right>=0.735\pm0.004(stat.)\pm0.146(syst.)$.
\footnote{
The weighted systematic error is obtained by calculating the weighted
average
over the five $Q_3$ bins separately for each kind of systematic error.
These errors are then added in quadrature.
On the other hand, adding quadratically the systematic errors of the five
$Q_3$
bins, as done for
weighted statistical errors,
would give
$\pm0.078$ instead of $\pm0.146$ for the systematic uncertainty.}

This result is in agreement with the previously published WA98 result
obtained using about half of the present
data sample \cite{C3WA98}, where a more detailed description of the
three-pion analysis method can also be found.
More recently, the NA44 experiment \cite{NA44bis} obtained with lower
statistics a factor $\left<W\right>$ which agrees with our results.

\section{Conclusions}

We have studied $m_T$ distributions for identified $\pi^-$ and $K^-$
produced in central Pb+Pb collisions at 158$A$ GeV.
The W.-H. hydrodynamical model has been fitted to the pion spectrum.
The resulting fitted parameters favor a combination of a relatively low
temperature
$T$ $\sim$85 MeV and an average transverse flow velocity
$\left<\beta_T\right>$ $\sim$0.50.
The shape of the $\pi^-$ $m_T$ distribution is in good agreement with
the $\pi^0$ $m_T$ distribution measured in the same experiment.

Bose-Einstein interferometry of $\pi^-$ pairs gives fitted radii of
typically 7 fm.
This has to be compared to the equivalent rms radius of the initial Pb ion
of 3.2 fm, indicating an expanded emission volume at freeze-out.

The analysis of two-pion correlations has been performed as a function
of $k_T$ using two different parameterizations in the LCMS.
The results are consistent between the standard 3-dimensional
Pratt-Bertsch fit and the Yano-Koonin-Podgoretski{\u{\i}} fit.

A clear dependence of all radius parameters on $k_T$ is observed,
with a stronger dependence for the longitudinal radii,
indicating a larger longitudinal than transverse expansion.
Both the cross term $R^2_{out-long}$ from the PB fit and $v_L$ from the
YKP fit deviate from zero, which suggests that the source seen within the
acceptance does not undergo a strictly boost invariant expansion.
Moreover the short duration of emission disfavours any long-lived
intermediate phase.

A comparison of the data with a hydrodynamical simulation based on
the Wiedemann-Heinz model, and taking
into account acceptance and resolution effects has been made for the radii
parameters as a function of $k_T$.
The agreement is found to be very good.

The average pion phase-space density at freeze-out has been
calculated from measured quantities as a function of $p_T$.
The results indicate a clear deviation from a Bose-Einstein
distribution at high $p_T$, but are very well fitted by the
Tom\'a{\u{s}}ik-Heinz model.
The pion chemical potential, which is included in the model, is found to
be compatible with zero,
while a large pion chemical potential close to the condensation
limit of $m_\pi$, seems to be excluded.

Finally, we have studied the $\pi^-\pi^-\pi^-$ interference and found a
substantial contribution of genuine three-pion correlations in central
collisions.
For $Q_3\leq$60 MeV/$c$ a weighted mean of the
strength of the genuine three-pion
correlations $\left<W\right>=0.735\pm0.004(stat.)\pm0.146(syst.)$ was
extracted.
This is somewhat smaller than what is expected for a fully chaotic and
symmetric source. 

\vspace{1 cm}
\vspace{5 mm}
\hspace{-5.5 mm}
{\normalsize{\textbf {Acknowledgements}}}
\vspace{3.5 mm}
\\
\hspace*{3 mm}We would like to thank the CERN-SPS
accelerator crew
for providing an excellent Pb beam. 
This work was supported jointly by the German BMBF and DFG, the U.S. 
DOE, the Swedish NFR, the Dutch Stichting FOM, the Swiss National Fund,
the Humboldt Foundation, the Stiftung f\"{u}r deutsch-polnische
Zusammenarbeit, the Department of Atomic Energy, the Department
of Science and Technology and the University Grants Commission of
the Government of India, the Indo-FRG Exchange Programme, the PPE
division of CERN, the INTAS under contract INTAS-97-0158, the
Polish KBN under the grant 2P03B16815, the Grant-in-Aid for Scientific
Research
(Specially Promoted Research \& International Scientific Research)
of the Ministry of Education, Science, Sports and Culture, JSPS
Research Fellowships for Young Scientists,
the University of Tsukuba Special Research Projects, and
ORISE.
ORNL is managed by UT-Battelle, LLC, for the U.S. Department of Energy
under contract DE-AC05-00OR22725.

\onecolumn
\begin{table*}[hbtp]
\caption{3-dimensional analysis as a function of $k_T$ for the PB
and the YKP fits.
$\lambda$ after correction for backgrounds from misidentified
pions is also shown.}

\label{tab:table1}
\begin{center}
\begin{tabular}{l|r@{$\pm$}lr@{$\pm$}lr@{$\pm$}lr@{$\pm$}lr@{$\pm$}l}
$\left<k_T\right>$&\multicolumn{2}{l}{\enspace 0.02 GeV/$c$}
&\multicolumn{2}{l}{\enspace 0.07 GeV/$c$}
&\multicolumn{2}{l}{\enspace 0.125 GeV/$c$}
&\multicolumn{2}{l}{\enspace 0.175 GeV/$c$}
&\multicolumn{2}{l}{\enspace 0.285 GeV/$c$}\\
\hline
$R_{TS}$&6.48&0.18 fm&6.67&0.15 fm&5.55&0.24 fm&5.13&0.32 fm&4.81&0.29
fm\\
$R_{TO}$&6.95&0.20 fm&6.62&0.16 fm&6.70&0.29 fm&6.29&0.43 fm&6.07&0.50
fm\\
$R_{L}$&8.69&0.26 fm&7.54&0.19 fm&5.95&0.27 fm&5.45&0.37 fm&5.15&0.34
fm\\
$R^2_{out-long}$&1.51&2.11 fm$^2$&0.55&1.45 fm$^2$&-0.52&1.96
fm$^2$&-3.01&2.31 fm$^2$&-4.51&1.94 fm$^2$\\
$\lambda$&0.334&0.012&0.362&0.010&0.337&0.018&0.324&0.026&0.391&0.036\\
$\lambda_{cor}$&0.646&0.023&0.634&0.018&0.557&0.030&0.511&0.041&0.596&0.055\\
$R_{T}$&6.66&0.14 fm&6.54&0.13 fm&5.99&0.19 fm&5.78&0.29 fm&5.17&0.26
fm\\
$R_{4}$&8.68&0.26 fm&7.52&0.19 fm&5.97&0.28 fm&5.49&0.39 fm&4.94&0.35
fm\\
$v_L$&0.02&0.08 fm&-0.04&0.06 fm&0.08&0.08 fm&0.14&0.10 fm&0.20&0.09
fm\\
\end{tabular}
\end{center}
\end{table*}
\begin{table*}[hbtp]
\caption{Average $\pi^-$ phase-space density $\left<f\right>$ at
freeze-out as a function of $p_T$.}
\label{tab:table2}
\begin{center}
\begin{tabular}{l|r@{$\pm$}lr@{$\pm$}lr@{$\pm$}lr@{$\pm$}lr@{$\pm$}l}
$\left<p_T\right>$&\multicolumn{2}{l}{\enspace 0.02 GeV/$c$}
&\multicolumn{2}{l}{\enspace 0.07 GeV/$c$}
&\multicolumn{2}{l}{\enspace 0.125 GeV/$c$}
&\multicolumn{2}{l}{\enspace 0.175 GeV/$c$}
&\multicolumn{2}{l}{\enspace 0.285 GeV/$c$}\\
\hline
$\left<f\right>$&0.319&0.018&0.272&0.012&0.237&0.019&0.182&0.022
&0.088&0.012\\
\end{tabular}
\end{center}
\end{table*}

\begin{table*}[hbtp]
\caption{Weighted mean of the strength of the genuine
three-pion correlations $\left<W\right>$ as a function of $Q_3$.}
\label{tab:table3}
\vspace*{0.8cm}
\hspace*{-2.0cm}
\begin{tabular}{l|rlrlrlrlrl}
$Q_3$ &\multicolumn{2}{l}{\enspace 0.01~-~0.02 GeV/$c$}
&\multicolumn{2}{l}{\enspace 0.02~-~0.03 GeV/$c$}
&\multicolumn{2}{l}{\enspace 0.03~-~0.04 GeV/$c$}
&\multicolumn{2}{l}{\enspace 0.04~-~0.05 GeV/$c$}
&\multicolumn{2}{l}{\enspace 0.05~-~0.06 GeV/$c$}\\
\hline
$\left<W\right>$&0.830&\begin{tabular}{r}$+0.201$\\$-0.175$\end{tabular}
&0.780&\begin{tabular}{r}$+0.156$\\$-0.143$\end{tabular}
&0.736&\begin{tabular}{r}$+0.145$\\$-0.138$\end{tabular}
&0.654&\begin{tabular}{r}$+0.177$\\$-0.193$\end{tabular}
&0.557&\begin{tabular}{r}$+0.258$\\$-0.293$\end{tabular}\\
\end{tabular}
\end{table*}

\end{document}